%% file: damagepaper_arxiv.tex
\documentclass[structabstract]{aa}  
\usepackage{graphicx}
\usepackage{txfonts}
\usepackage{units}
\usepackage{natbib} \bibpunct{(}{)}{,}{a}{}{,}
\usepackage{upgreek}
\usepackage{color}
\include{definedquantities}

\begin{document}
   \title{Collisions of inhomogeneous pre-planetesimals}

   \author{R.J. Geretshauser\inst{1}
   	\and
          R. Speith\inst{2}
          \and
          W. Kley\inst{1}
          }

   \institute{Institut f\"ur Astronomie und Astrophysik, Abteilung Computational Physics,
   	    Eberhard Karls Universit\"at T\"ubingen, Auf der Morgenstelle 10, 72076
	    T\"ubingen, Germany\\
	     \email{ralf.j.geretshauser@uni-tuebingen.de}
	     \and
	    Physikalisches Institut, Eberhard Karls Universit\"at T\"ubingen, Auf der 	
   	    Morgenstelle 14, 72076 T\"ubingen, Germany
            }

   \date{}

 
  \abstract
   {In the framework of the coagulation scenario, kilometre-sized planetesimals form by subsequent collisions of pre-planetesimals of sizes from centimetre to hundreds of metres. Pre-planetesimals are fluffy, porous dust aggregates, which are inhomogeneous owing to their collisional history. Planetesimal growth can be prevented by catastrophic disruption in pre-planetesimal collisions above the destruction velocity threshold.}
   {We assess whether the inhomogeneity created by subsequent collisions has a significant influence on the stability of pre-planetesimal material to withstand catastrophic disruption. We wish develop a model to explicitly resolve inhomogeneity. The input parameters of this model must be easily accessible from laboratory measurements.}
{We develop an inhomogeneity model based on the density distribution of dust aggregates, which is assumed to be a Gaussian distribution with a well-defined standard deviation. As a second input parameter, we consider the typical size of an inhomogeneous clump. For the simulation of the dust aggregates, we utilise a smoothed particle hydrodynamics (SPH) code with extensions for modelling porous solid bodies. The porosity model was previously calibrated for the simulation of \sio dust, which commonly serves as an analogue for pre-planetesimal material. The inhomogeneity is imposed as an initial condition on the SPH particle distribution. We carry out collisions of centimetre-sized dust aggregates of intermediate porosity. We vary the standard deviation of the inhomogeneous distribution at fixed typical clump size. The collision outcome is categorised according to the four-population model.}
   {We show that inhomogeneous pre-planetesimals are more prone to destruction than homogeneous aggregates. Even slight inhomogeneities can lower the threshold for catastrophic disruption. For a fixed collision velocity, the sizes of the fragments decrease with increasing inhomogeneity.}
   {Pre-planetesimals with an active collisional history tend to be weaker. This is a possible obstacle to collisional growth and needs to be taken into account in future studies of the coagulation scenario.}


   \maketitle
%

\section{Introduction}

Terrestrial planets are thought to form by core accretion in protoplanetary accretion discs \citep[e.g.][]{Lissauer.1993}. The discs consist of both hydrogen and helium gas and a small fraction of dust, which in the early stages exists as micron sized dust grains \citep[e.g.][]{Dullemond.2007}. Interaction between the gas and dust induces relative motions between the dust grains and causes collisions among them \citep[e.g.][]{Weidenschilling.1993}. Mutual sticking by van der Waals forces occurs and by this process millimetre- to centimetre-sized, highly porous pre-planetesimals form \citep[e.g.][]{Blum.2008}. Additional growth by means of collision is endangered by several obstacles: rebound, material loss by accretion or photoevaporation, and fragmentation. \citet{Zsom.2010} found that the growth of pre-planetesimals to kilometre-sized planetesimals may be halted by mutual compaction and bouncing, which leads finally to a population of dust and pebble-sized objects in the protoplanetary disc. \citet{Weidenschilling.1977} found that metre-sized bodies have the highest drift speed towards the host star, where the objects get accreted or photoevaporated. For these objects, the drift timescale is much shorter than the time needed for a planet to form. One of the most serious obstacles to planetesimal formation is the fragmentation barrier: with increasing aggregate size the mutual collision velocities also increase and catastrophic disruption might become more common than mass gain \citep[e.g.][]{Brauer.2008}.

If by some mechanism a sufficient population of kilometre-sized objects is formed, their self-gravity acts as an accretion mechanism and ensures further growth to planets. In this gravity dominated regime, kilometre-sized objects and yet larger ones are referred to as planetesimals.

To investigate planetesimal formation, global dust coagulation models \citep[e.g.][]{Dullemond.2005, Zsom.2010, Birnstiel.2010} numerically compute the evolution of dust aggregates in a protoplanetary disc starting from micron-sized dust grains. These models require detailed collision statistics evaluated for the outcome of two-body pre-planetesimal collisions as functions of important parameters such as mass ratio, porosity, impact parameter, and relative velocity of the collision partners. In the past decade, many experimental studies have been carried out to collect these data \citep[e.g.][for summaries]{Blum.2008, Guttler.2010}. However, laboratory setups become difficult and infeasible for centimetre-sized pre-planetesimals and larger ones since they have to be carried out under protoplanetary disc conditions, i.e.\ in a vacuum and under micro-gravity. In this regime, collision data has to be acquired by numerical simulations.

In their pioneering work, \citet{Benz.1994,Benz.1995} developed a solid body smoothed particle hydrodynamics (SPH) implementation for brittle material based on the dynamical fracture model by \citet{Grady.1980}, which evolves the propagation of cracks starting from an initial \citet{Weibull.1939} distribution of flaws. On these grounds, \citet{Benz.2000} investigated collisions between rocky non-porous pre-planetesimals in their typical collision velocity regime from $\unit[5]{\msec}$ to $\unit[40]{\msec}$ and found only disruption. However, experiments \citep[e.g.][]	{Blum.2008} indicate that pre-planetesimals are fluffy, porous objects rather than solid rocks. For this reason, \citet{Sirono.2004} developed an SPH implementation of a porosity model based on porosity-dependent material quantities such as bulk and shear moduli as well as compressive, shear, and tensile strengths. In his simulations of porous ice aggregates, he found that owing to their porous structure the kinetic energy can be dissipated effectively, permitting collisional growth. Sirono's approach included a damage model based again on the dynamical fracture model of \citet{Grady.1980}. In addition, a damage restoration model tries to capture damage healing by compression. However, \citet{Schafer.2007} found that Sirono's damage model is not applicable to their SPH simulations of porous ice because it includes damage by compression. In contrast, materials such as porous ice and porous dust form new molecular bondings, when they are compressed. Thus, their strengths increase but do not decrease as predicted by the Sirono damage model. On the basis of the implementation of \citet{Benz.1994,Benz.1995}, \citet{Jutzi.2008} combined the damage model of \citet{Grady.1980} and the $p-\alpha$ model of \citet{Herrmann.1969} to simulate porous brittle media. The model was calibrated with laboratory high velocity impact experiments \citep{Jutzi.2009}. However, the model by \citet{Jutzi.2008} is also not applicable to porous protoplanetary dust because it includes damage by compression and has no suitable model of damage healing.

To simulate \sio dust, which is used as protoplanetary dust analogue in many experimental studies, we developed a suitable porosity model based on the ideas of \citet{Sirono.2004}, that was implemented into a solid body SPH code to numerically investigate pre-planetesimal collisions \citep{Geretshauser.2010}. With the aid of laboratory benchmark experiments \citep[see also][]{Guttler.2009}, the porosity model was calibrated and tested for the simulation of \sio dust. In particular, the simulation reproduced the compaction, bouncing, and fragmentation behaviour of homogeneous dust aggregates of various porosities in a quantitatively correct way. However, this approach does not include an explicit damage model.

In this paper, we present an approach to explicitly resolve inhomogeneity of dust aggregates. Since an inhomogeneous density distribution is one of the main sources of fracture in porous aggregates, this work mimics a damage evolution by the evolution of the density. The inhomogeneity is described by a Gaussian density distribution and the typical size of a clump. The aggregate inhomogeneity of \sio dust can easily be measured in the laboratory by X-ray tomography \citep[see][Fig.~10]{Guttler.2009}. In contrast, the \citet{Grady.1980} damage model relies on the correct parameters for the initial Weibull distribution of flaws, which are difficult to determine in laboratory experiments. Our approach of resolving inhomogeneity is closely connected to the porosity model, where the strength quantities are porosity dependent. Therefore, any fluctuation in porosity induces a fluctuation in strength over the aggregate, which is utilised to model the evolution of the material under stress.

To classify the collisional outcome of our simulations involving inhomogeneous aggregates, we utilise the four-population model. This classification scheme was introduced by \citet{Geretshauser.2011} to enable a precise mapping of all collisional outcomes involving any combination of sticking, bouncing, and fragmentation processes. Four fragment populations are distinguished according to their masses: both the largest and the second largest fragment as well as a set of fragments whose mass distribution can be approximated by a power law. To take into account the resolution limit, the sub-resolution population contains all fragments that consist of a single SPH particle.

The outline of this paper is the following. In Sect.\ \ref{sec:porosity-model}, we describe the SPH method with extensions for the simulation of solid bodies and the applied porosity model. Section \ref{sec:inhomogeneity-damage-model} firstly describes a previous study of a damage model, then  we present our approach based on the inhomogeneity of dust aggregates and its implementation. The results of collisions with inhomogeneous aggregates are discussed in Sect.\ \ref{sec:results}. We summarise our findings in Sect.\ \ref{sec:discussion}, where we also discuss the results and provide an outlook on future work.


\section{Solid body SPH and porosity model}
\label{sec:porosity-model}

For our simulations, we apply the smoothed particle hydrodynamics (SPH) numerical method originally developed by \citet{Lucy.1977} and \citet{Gingold.1977} for astrophysical flows. In subsequent studies, SPH was extended to model the elastic and plastic deformations of solid materials \citep[e.g.][]{Libersky.1991, Benz.1994, Randles.1996}. Our parallel SPH code \texttt{parasph} \citep{Hipp.2004, Schafer.2005, Schafer.2007, Geretshauser.2010} is based on these concepts. In the SPH scheme, the time evolution of a solid body is simulated by means of the Lagrangian equations of continuum dynamics
\begin{eqnarray}
\totder{\rho}{t} & = &- \rho \partder{\va}{\xa} \; , \\
\totder{\va}{t} & = & \frac{1}{\rho} \partder{\stab}{\xb} \; ,
\end{eqnarray}
which represent the continuum and momentum equations, respectively. The quantities $\rho$, $\va$, and $\stab$ have their usual meanings, i.e.\ density, velocity, and the stress tensor. The Greek indices denote the spatial components and the Einstein sum convention applies to them. Since our equation of state (EOS) is energy-independent, we do not solve the energy equation, hence omit to state it here. Within the framework of SPH, the continuous quantities are discretised into mass packages, which we refer to as ``particles''. These represent the sampling points of the method and interact with each other within a limited spatial range, called the smoothing length $h$. The influence of a particle $b$ on a particle $a$ depends on the particle distance $|\vx^a - \vx^b|$ and is weighted by the kernel function $W(|\vx^a - \vx^b|, h)$. This is usually a spherically symmetric function with compact support, which is differentiable at least to first order. We utilise the cubic spline kernel of \citet{Monaghan.1985}.

The particles of the SPH scheme move according to the Lagrangian form of the equations of continuum mechanics. Thus, they represent a natural frame of reference for any deformation of the simulated solid body. In the context of fragmentation, two fragments are considered to be separate once their constituent subsets of particles no longer interact, i.e., when the fragments are separated by more than $2\,h$. The fragments may be spread out over an unlimited computational domain. This is why computations are carried out at the particle positions.

The stress tensor in the momentum equation can be divided into the term accounting for pure hydrostatic pressure $p$ and the term representing pure shear, which is given by the traceless deviatoric stress tensor $\Sab$. Hence,
\begin{equation}
\stab = - p \dab + \Sab \; .
\label{eq:stress-tensor}
\end{equation}
In contrast to viscous fluids, $p$ and $\Sab$ do not depend on the velocity gradient but on the deformation of a solid body given by the strain tensor
\begin{equation}
\epsilon^{\alpha\beta} = \frac{1}{2} \left(
	\frac{\partial {x'}^\alpha}{\partial x^\beta}
	+ \frac{\partial {x'}^\beta}{\partial x^\alpha}
	\right) \, ,
\end{equation}
where $\vx'$ are the coordinates of the deformed body. In particular, the relations that hold are
\begin{eqnarray}
p  & = & - K \epsilon_{\alpha\alpha} \label{eq:elastic-pressure} \, ,\\
S^{\alpha\beta} & = & 2\mu \left(
	\epsilon^{\alpha\beta} - \frac{1}{d} \delta^{\alpha\beta} 
        \epsilon^{\gamma\gamma}
	\right) + \mathrm{rotation~terms},
	\label{eq:elastic-shear}
\label{eq:elastic-deviatoric}
\end{eqnarray}  
where the quantity $d$ denotes the dimension and $K$ and $\mu$ are the bulk modulus and shear modulus, respectively, for elastic deformation. To obey the principle of frame invariance, rotation terms have to be added to the deviatoric stress tensor. For this, we apply the Jaumann rate form, which was previously used by \citet{Benz.1994} and \citet{Schafer.2007}, and described by \citet{Geretshauser.2010} in more detail.

The above relations (Eqs. \ref{eq:elastic-pressure} and \ref{eq:elastic-shear}) describe the time evolution of a solid body in the elastic regime, i.e., for reversible deformations. With respect to pre-planetesimal collisions, we are concerned with highly porous material that undergoes elastic and plastic deformations. Both aspects are described by means of the porosity model of \citet{Geretshauser.2010}, which is based on the ideas of \citet{Sirono.2004} and can be categorised as a plasticity-based porosity model. 

In this approach, the quantities for the elastic regime (bulk and shear moduli), as well as the strength quantities (compressive, shear, and tensile strength), depend on the filling factor, which accounts for the sub-resolution porosity. The filling factor $\phi$ is related to the density $\rho$ and the porosity $\Phi$ by
\begin{equation}
\phi = 1 - \Phi = \frac{\rho}{\rhos} \; ,
\end{equation}
where $\rhos$ is the density of the matrix material. We use a constant value of $\rhos = \unit[2000]{kg\,m^{-3}}$ for our \sio material \citep[e.g.][]{Blum.2004}. In particular, the elastic quantities are related to the filling factor by a power law
\begin{equation}
K (\phi) = 2 \mu(\phi) = K_0 \left( \frac{\phi}{\phirbd} \right)^\gamma \; ,
\label{eq:bulk-modulus}
\end{equation}
where $\gamma = 4$, $K_0$ is the bulk modulus of an uncompressed random ballistic deposition (RBD) dust sample with $\phirbd = 0.15$ \citep[e.g.][]{Blum.2004}, such that $K_0 = K(\phirbd) = \unit[4.5]{kPa}$.

For the compressive, tensile, and shear strength we adopt some empirical relations (see \citealt{Guttler.2009}, \citealt{Geretshauser.2010} and \citealt{Geretshauser.2011} for details). These relations allow us to reproduce the compaction, bouncing, and fragmentation behaviour to high accuracy in a velocity range of the order of $0.1$ to $\unit[10]{\msec}$ \citep{Geretshauser.2010}. \citet{Geretshauser.2011} showed that all collision types found in laboratory experiments can be reproduced. In the same reference, the laboratory velocity thresholds for the transitions between sticking, bouncing, and fragmentation behaviour resembled the simulation results for collisions of medium porosity aggregates with up to $\unit[27.5]{\msec}$. Therefore, we expect these relations to be valid for collisions below the sound speed of the dust material, which is $\sim \unit[30]{\msec}$ \citep{Guttler.2009}. For the compressive strength,
\begin{equation}
\Sigma(\phi) = \pmean \left( \frac{\phi_{\rm max} - \phi_{\rm min}}{\phi_{\rm max} - \phi} - 1 \right)^{\Delta\,\ln{10}} \; ,
\label{eq:compressive-strength}
\end{equation}
where $\phi_{\rm min} + \varepsilon < \phi < \phi_{\rm max}$ and $\varepsilon = 0.005$. The quantities $\phi_{\rm min} = 0.12$ and $\phi_{\rm max} = 0.58$ are the minimum and maximum filling factors, respectively, in the compressive strength relation. Both values can be exceeded by the material. The power of the compressive strength relation is $\ln(10)$ times the parameter $\Delta$ with $\Delta = 0.58$. The quantity $\pmean = \unit[260]{Pa}$ accounts for the mean pressure of the compressive strength relation. According to \citet{Geretshauser.2010, Geretshauser.2011}, these parameters are valid for impacts below $\sim \unit[30]{\msec}$. The influence of $\Delta$ and $\pmean$ was studied by \citet{Geretshauser.2010}. For $\phi \le \phi_{\rm min} + \varepsilon$, the compressive strength relation is continuously extended by the constant function $\Sigma(\phi) = \Sigma(\phi_{\rm min} + \varepsilon)$ and for $\phi_{\rm max} \le \phi$, we set $\Sigma(\phi) = \infty$. The tensile strength is given by
\begin{equation}
T(\phi) = \unit[- 10^{2.8 + 1.48\phi}]{Pa} \;
\end{equation}
and the shear strength is the geometric mean of the compressive and tensile strength
\begin{equation}
Y(\phi) = \sqrt{\Sigma(\phi) | T(\phi) |} \; .
\end{equation}
We note that the three strength quantities are filling-factor dependent. This is because as the filling factor increases the monomers of the dust material are packed more closely. Thus, each monomer establishes more bonds, which increases the strength. The filling-factor dependence of the strength quantities is exploited by the approach presented in this paper. Together the bulk modulus, the compressive strength, and the shear strength yield the full equation of state for pure hydrostatic compression or tension
\begin{equation}
p (\phi) = \left\{
\begin{array}{ll}
\Sigma(\phi) & \mathrm{for}~\phi_c^+ < \phi \\
K(\phi_0')(\phi/\phi_0' - 1) & \mathrm{for}~\phi_c^- \le \phi \le \phi_c^+\\
T(\phi) & \mathrm{for}~\phi < \phi_c^-
\end{array} \right. \; ,
\label{eq:porosity-model}
\end{equation}
where $\phi_c^+ > \phi_c^-$, and $\phi_c^+$ and $\phi_c^-$ are critical filling factors. The value of $\phi_c^+$ marks the transition between elastic and plastic compression and $\phi_c^-$ defines the transition between elastic and plastic tension. The filling factors in-between the critical values represent the elastic regime. The quantity $\phi_0'$ is the reference filling factor that represents the filling factor of the porous material at vanishing external pressure. Once the material undergoes plastic deformation, the reference filling factor is reset using the middle relation of Eq.\ \ref{eq:porosity-model}. Details of this pressure reduction process can be found in \citet{Geretshauser.2010}. In the elastic regime (middle line of Eq.\ \ref{eq:porosity-model}), the pressure varies with density according to the elastic path given by Eq.\ \ref{eq:bulk-modulus}. If $\phi_c^+$ is exceeded, the pressure follows the much shallower compressive strength path $\Sigma(\phi)$ in the positive $\phi$-direction, whereas if $\phi_c^-$ is exceeded, the pressure follows the shallower tensile strength path $T(\phi)$ in the negative $\phi$-direction \citep[cf.\ Fig.\ 1 in][]{Geretshauser.2010}. In both cases, the absolute value of the pressure, which is computed according to the elastic equation of state, is reduced to the compressive or tensile strength, respectively. This accounts for the stress relieved by plastic deformation. Details of the implementation are described by \citet{Geretshauser.2010}.

For the plastic deformation by pure shear, the deviatoric stress tensor $\Sab$ is reduced according to the von Mises plasticity given by \citep[e.g.][]{Benz.1994}
\begin{equation}
\Sab \rightarrow f \Sab \; .
\end{equation}
The function $f$ is defined by $f = \min \left[ Y^2 / 3 J_2, 1 \right]$. An alternative approach \citep[e.g.][]{Wilkins.1964, Libersky.1993} uses the expression $f = \min \left[ \sqrt{Y^2 / 3 J_2}, 1 \right]$ but we expect this to lead to differences from our adopted approach that are marginal. In this relation, $J_2 = 0.5 \, \Sab \Sab$ and $Y = Y(\phi)$ are the second irreducible invariant of the deviatoric stress tensor and the shear strength, respectively.


\section{Resolving inhomogeneity}
\label{sec:inhomogeneity-damage-model}

\subsection{Previous approach}
\label{subsec:previous-approaches}

In contrast to ductile media, brittle materials, such as basalt, granite, or porous pumice, do not rupture by plastic flow. This is because the material is not completely homogeneous but contains little flaws. These are little defects in the medium. With increasing strain, cracks develop originating from these flaws and start to pervade the solid body. In brittle media, stress is relieved by developing cracks.

To model damage, \citet{Grady.1980} introduce a scalar parameter $0 \le D \le 1$, where $D= 0$ represents undamaged and $D = 1$ fully disintegrated material. The stress tensor (Eq.\ \ref{eq:stress-tensor}) is modified according to
\begin{equation}
\sigma_{\rm dam} = (1 - D)\sigma \; .
\label{eq:damage-stress}
\end{equation}
Thus, damaged material may feel less stress than undamaged material. The local damage $D$ is defined as the ratio of the volume defined by the growing crack to the volume in which the crack is growing \citep[see][]{Grady.1980}. The local time evolution of the damage $D$ is given by 
 \begin{equation}
 \totder{D^{\nicefrac{1}{3}}}{t} = n_{\rm act} \frac{c_{\rm g}}{R_{\rm s}} \; ,
 \end{equation}
where $n_{\rm act}$ accounts for crack accumulation and denotes the number of activated flaws, and $R_{\rm s}$ is the radius of a circumscribing sphere in which the crack evolves. It accounts for the maximum size of a damaged area. The quantity $c_{\rm g}$ is the crack growth velocity. The concept of flaw activation originates from the idea that cracks do not start to grow for any strain applied but are activated for some strain threshold $\epsilon_{\rm act}$. The number of flaws $n$ per unit volume with $\epsilon_{\rm act}$ is given mostly by a power law
\begin{equation}
n(\epsilon_{\rm act}) = k_{\rm wb}\epsilon_{\rm act}^{m_{\rm wb}} \; .
\label{eq:weibull-distribution}
\end{equation}
This distribution of flaws in a brittle material was proposed by \citet{Weibull.1939}. It is based on two material parameters $k_{\rm wb}$ and $m_{\rm wb}$, where $k_{\rm wb}$ is the number of flaws per unit volume. The flaw distribution according to Eq.~\ref{eq:weibull-distribution} is set as an initial condition for the material before the simulation.

A disadvantage is that the material parameters $k_{\rm wb}$ and $m_{\rm wb}$ are rarely available because they are difficult to measure. Unfortunately, small variations in $k_{\rm wb}$ and $m_{\rm wb}$ lead to large differences in the activation thresholds and simulation results \citep[e.g.][]{Schafer.2005, Jutzi.2009}.

Because the damage model by \citet{Grady.1980} is only suitable for brittle material, it is not applicable to pre-planetesimal material represented by $\rm SiO_2$ dust, which has properties that are between those of a ductile and a brittle material. Therefore, it is desirable to develop a simple damage approximation consistent with the applied porosity model. To summarise, three ingredients are essential for a damage model: (1) a distribution of defects, (2) the stress reduction due to damage, and (3) the typical size of a damaged area.

\subsection{The approach of resolving inhomogeneity}

In contrast to the damage model presented in Sec.\ \ref{subsec:previous-approaches}, our approach does not explicitly consider the evolution of defects, hence is not strictly a damage model. However, it is designed to behave as an approximation to a damage model in some aspects, which are described in this section. As a significant advantage, our approach relies on quantities that that can be more easily measured by laboratory experiments than the parameters of the model by \citet{Grady.1980}.

The starting point for our approach is the inhomogeneous nature of the material. Macroscopic dust aggregates created by the RBD method are not completely homogeneous \citep{Guttler.2009}. The filling factor instead was found to follow a Gaussian distribution around a median of $\phi_\mu \sim 0.15$ with $\phis = 0.013$. 

According to the porosity model of Sec.~\ref{sec:porosity-model}, regions of lower filling factor also represent regions of weaker compressive, tensile, and shear strengths, whereas regions of higher filling factor are stronger. Therefore, as an analogue to the Weibull distribution we propose an initial distribution of the filling factor given by the Gaussian function
\begin{equation}
n(\phi) = \frac{1}{\sqrt{2 \uppi} \phi_{\sigma}} 
\exp{\left( - \frac{1}{2} \frac{\phi - \phi_{\mu}}{\phi_{\sigma}}\right)} ,
\label{eq:gaussian-distribution}
\end{equation}
where $n(\phi)$ is the number density for a filling factor $\phi$, $\phi_\sigma$ is the width of the distribution function, and $\phi_\mu$ is the median filling factor. The pressure limits of compressive, tensile, and shear strengths $\Sigma(\phi)$, $T(\phi)$, and $Y(\phi)$, respectively, mark the transition from the elastic to the plastic hydrostatic regime. Therefore, they can be regarded as analogues to the activation threshold $\epsilon_{\rm act}$. Hence, we can associate with each $\phi$ analogues to the activation threshold, which accounts for the first criterion of a damage model (see end of Sect.\ \ref{subsec:previous-approaches}) that is based on the distribution of defects.

Damaged regions in this simple inhomogeneity scheme are therefore represented by areas of low filling factor. Under constant tension, $\phi$ becomes increasingly smaller with time in these regions and, owing to the $\phi$-dependence of either the tensile strength $T(\phi)$ or shear strength $Y(\phi)$, the strength decreases in these regions as in Eq.~\ref{eq:damage-stress} with increasing $D$. This fulfils the second criterion: the stress reduction.

Instead of artificially introducing a damage parameter $D$ and an activation threshold $\epsilon_{\rm act}$, the filling factor $\phi$ naturally takes over this twofold role. On the one hand, it represents a quantity which determines the activation thresholds for the plastic regime, and on the other, it also represents a damage parameter that reduces, which decreases the strength quantities. Therefore, in the inhomogeneity approach, the time evolution of the approximated damage is given by the time evolution of $\phi$ or equivalently the density. The propagation speed of the defects is, intrinsically, the sound speed.

We introduce a typical size $R_{\rm c}$ of an inhomogeneous clump to comply with the third criterion. In terms of this parameter, an absolute length scale is introduced into the inhomogeneity approach. 

A great advantage of our approach is that the material parameters $\phi_\sigma$, $\phi_\mu$, and $R_{\rm c}$ can be determined in laboratory measurements, e.g.\ by X-ray tomography \citep{Guttler.2009}. However, no systematic study has been performed yet. A single measurement for a random ballistic deposition dust sample, which has the lowest filling factor ($\phi = 0.15$) and the highest degree of homogeneity, has found a standard deviation of $\phi_\sigma \sim 0.013$ \citep[Figure 10 in][]{Guttler.2009}. There exist no data on either the typical clump size or clump size distribution. Therefore, as a first approximation, in Sect.\ \ref{sec:results} we study the effect of $\phi_\sigma$ on the outcome of pre-planetesimal collisions for a fixed clump size $R_{\rm c}$. The influence of the latter will be studied in future work.

\subsection{Implementation issues}

\begin{figure}
   \resizebox{\hsize}{!}
            {\includegraphics{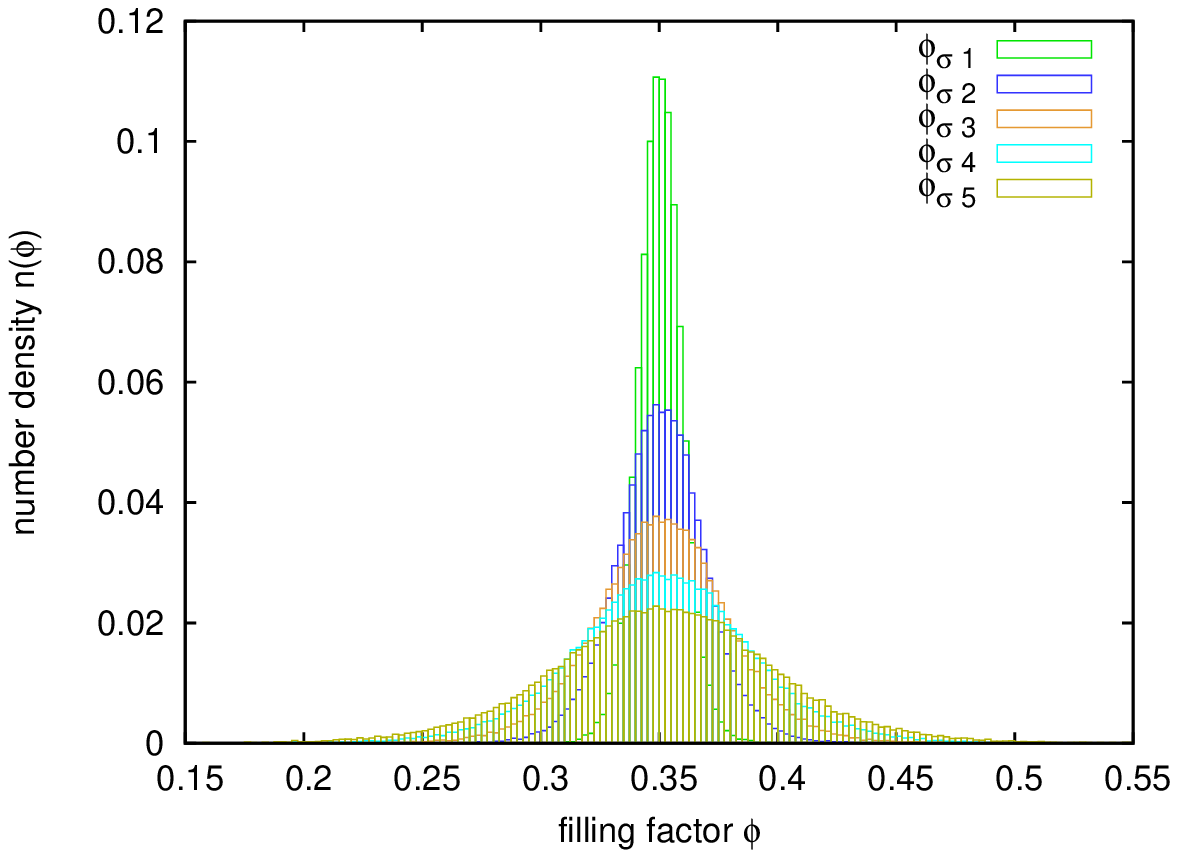}}
   \resizebox{\hsize}{!}
            {\includegraphics{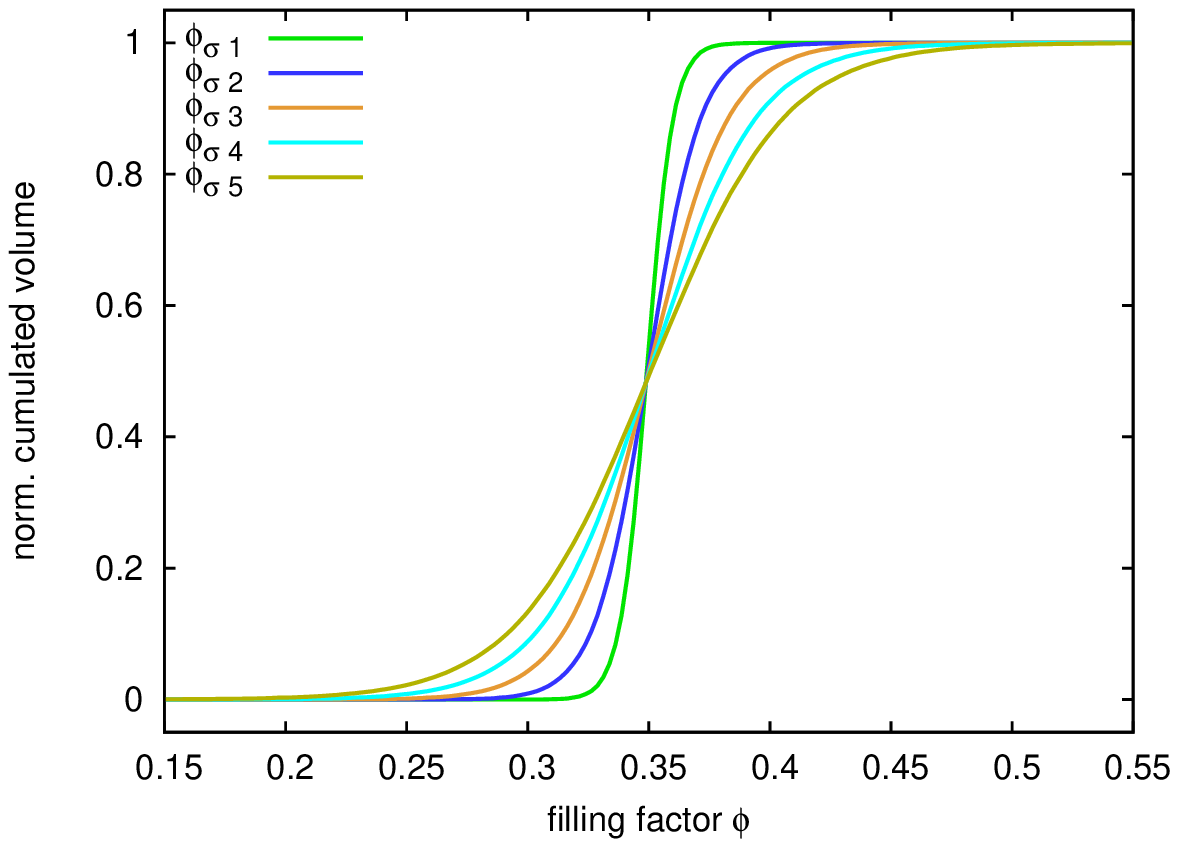}}
      \caption{Initial filling factor distributions of the target and projectile as binned (top) and cumulative diagram (bottom). The number density $n(\phi)$ of volumes with filling factor $\phi$ is plotted over $\phi$. A larger degree of inhomogeneity is characterised by the increasing standard deviation $\phis$ of the Gaussian (Eq.~\ref{eq:gaussian-distribution}) around the initial filling factor $\phii$ of the homogeneous aggregate. Here, we choose $\phii = 0.35$. The values of $\phi_{\sigma i}$ can be found in Table \ref{tab:phis}.}
         \label{fig:initial-ff}
\end{figure}

\begin{figure*}
   \centering
   \resizebox{0.9\hsize}{!}
            {\includegraphics{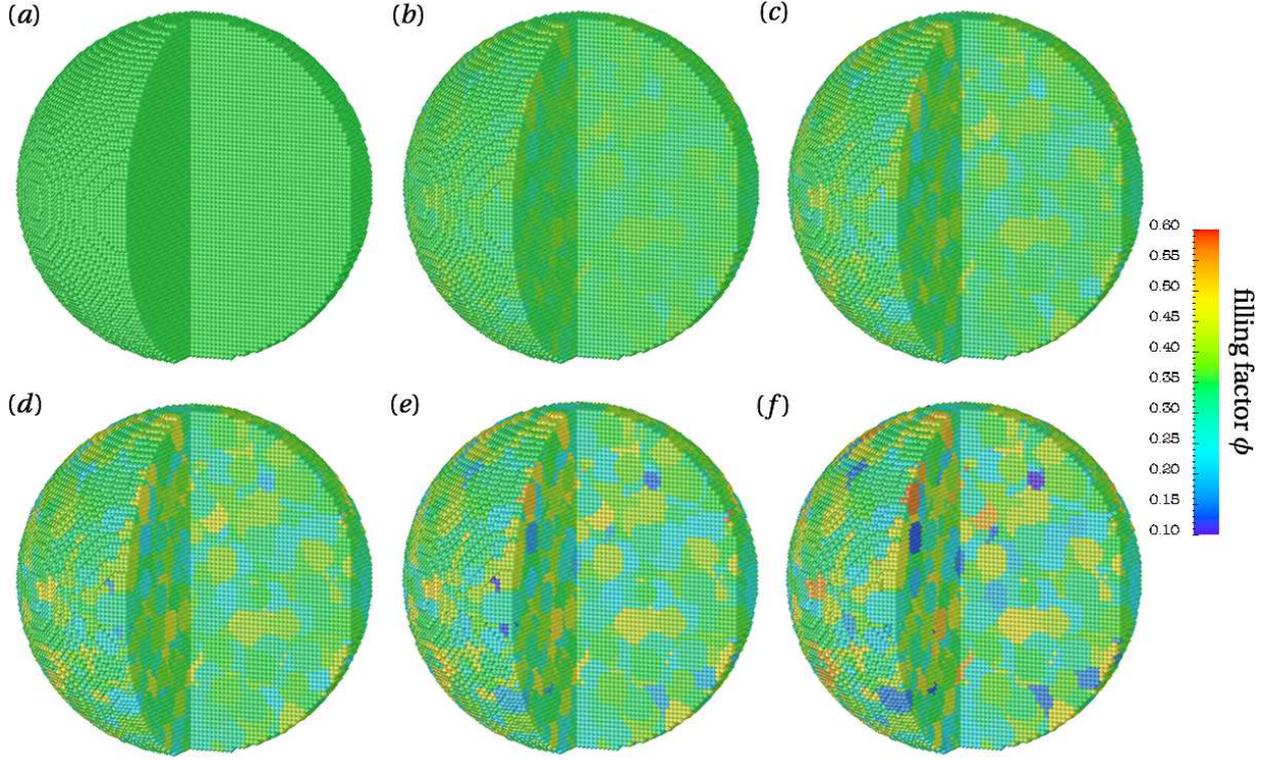}}
      \caption{Interior and exterior view of the targets for different inhomogeneities with the filling factor colour coded. The inhomogeneity increases from the homogeneous target (a) to the most inhomogeneous target (f). In particular, the standard deviations are $\phis = 0$ (a), $\phis = 0.01$ (b), $\phis= 0.02$ (c), $\phis = 0.03$ (d), $\phis = 0.04$ (e), and $\phis = 0.05$ (f). All targets have a similar pattern that originates from considering interacting particle passages in the implementation. With increasing $\phis$ the maximum and minimum filling factor of the different spots increase and decrease, respectively. The projectiles have a similar appearance. The result of collisions among these aggregates with $\unit[10]{\msec}$ are depicted in Fig.~\ref{fig:final-state}.}
         \label{fig:initial-setup}
\end{figure*}

The inhomogeneity is imposed on the simulated aggregate as an initial condition. Initially, we generate a homogeneous aggregate by assigning a constant initial filling factor $\phii$. As a second step, $\phii$ is modified according to a Gaussian distribution with a standard deviation $\phis$, which we will simply call either the ``inhomogeneity'' or ``degree of inhomogeneity'' in this paper. For the distribution, we randomly select a particle $a$ and set its density to a new $\tilde{\phii}^a$ following the Gaussian. The same $\tilde{\phii}^a$ is assigned to all particles within a radius of typical clump, which is fixed to $R_{\rm c} = \unit[9.75]{mm}$, the equivalent of one smoothing length $h$. Together with a spatial distance of \unit[2.6]{mm} between the SPH particles, one clump is resolved by $\sim 290$ particles, which is well above the required resolution \citep{Geretshauser.2010}. Since there are no empirical data for the typical clump size yet, as a first approximation we assume that the clump size is constant. With the chosen size, the simulated spherical decimetre bodies can be regarded as being assembled from roughly spherical centimetre aggregates in sticking processes with different velocities similar to those described by \citet{Guttler.2010} and \citet{Geretshauser.2011}. However, the effect of the typical clump size and resolution on the resulting fragment distribution will be studied in future work.

It is very likely that dust aggregates are compacted during their evolution in the protoplanetary disc \citep{Zsom.2010}, in particular if the aggregates are assembled by smaller units in sticking processes. For this reason, we do not consider very fluffy aggregates with $\phii = 0.15$ as produced by random ballistic deposition. We instead simulate spheres whose filling factor lies between the minimum ($\phimin = 0.12$) and maximum ($\phimax = 0.58$) filling factor of the compressive strength relation (Eq.\ \ref{eq:compressive-strength}), i.e. $\phii = 0.35$.

The mass of all individual SPH particles, which is a numerical parameter in the SPH equations, takes a constant value in the simulation. This value was computed to be consistent with the initial homogeneous aggregate (here $\phii = 0.35$). Therefore, the modification of the density introduces a slight inconsistency in the SPH particle distribution. This means that the density value of an arbitrary particle differs slightly from the value computed from the particle number density using the standard SPH sum approach. Potentially, this could introduce pressure fluctuations causing spurious particle motion. However, in our algorithm the density is determined by solving the continuity equation and not by evaluating the number density of the particles \citep[e.g.][]{Monaghan.2005}. As a consequence this inconsistency is marginal. Moreover, the time evolution of a inhomogeneous dust aggregate at rest was simulated and no spurious particle motion has been detected. In addition, the aggregate displayed no signs of instabilities within simulation times much longer than the collision timescale. Consequent	ly, the presented approach is assumed to yield stable aggregates. Evaluating the generated SPH particle distribution yields the resulting filling factor distribution: the number density $n(\phi)$ of the volume fraction with filling factor $\phi$ is  shown in Fig.~\ref{fig:initial-ff} as a binned and cumulative plot for various $\phis$. 

The inhomogeneous clumps with different filling factors are clearly visible in Fig.~\ref{fig:initial-setup}, where the density structure of the target is colour coded for different $\phis$. The packages are picked randomly but the random pattern is the same for all aggregates to ensure that the results are comparable. Owing to the increasing $\phis$, the regions with $\phi \ne \phii$ gain higher maximum and lower minimum filling factors. Consequently, the inhomogeneity pattern for $\phi_{\sigma 1}$ and $\phi_{\sigma 2}$ (see Tab.~\ref{tab:phis}) is barely visible in Fig.~\ref{fig:initial-setup} (first row, second and third). In contrast, for $\phi_{\sigma 5}$ the difference between the extreme values of the filling factor create a clearly visible pattern with high contrast. 

The standard deviation of the Gaussian filling factor curve $\phis$ is supplied to the distribution routine as an input parameter. The range of $\phis$ begins slightly below the empirical value for a homogeneous random ballistic deposition aggregate ($\phis = 0.013$). There is no empirical data on the inhomogeneity of aggregates resulting from subsequent collisions. However, owing to the collisional history of an aggregate with macroscopic pre-planetesimals it is very likely that realistic aggregates have a higher degree of inhomogeneity. Therefore, we increase $\phis$ in steps of 0.01 up to $\phis = 0.050$. The real standard deviation computed from the full width at half maximum (FWHM) is smaller than the $\phis$ supplied to the SPH particle initialisation algorithm (algorithmic value). The FWHM values are listed in Tab.~\ref{tab:phis}. For the sake of simplicity, we use the algorithmic value this paper.

\begin{table}
\centering	
\begin{tabular*}{\hsize}{l@{\extracolsep{\fill}}c c}
\hline\hline
Symbol & algorithmic value & FWHM value \\
\hline
\\
$\phi_{\sigma 1}$	&	0.010	& 0.0085 \\
$\phi_{\sigma 2}$	&	0.020	& 0.018 \\
$\phi_{\sigma 3}$	&	0.030	& 0.027 \\
$\phi_{\sigma 4}$	&	0.040	& 0.034 \\
$\phi_{\sigma 5}$	&	0.050	& 0.042 \\
\hline
\end{tabular*}
\caption{Selection of the standard deviations $\phis$ as they are used in the inhomogeneity algorithm. They are compared with the achieved value deduced from the FWHM of resulting filling factor distribution in Fig.~\ref{fig:initial-ff}.}
\label{tab:phis}
\end{table}


\section{Simulation results}
\label{sec:results}

\begin{figure*}
   \resizebox{\hsize}{!}
            {\includegraphics{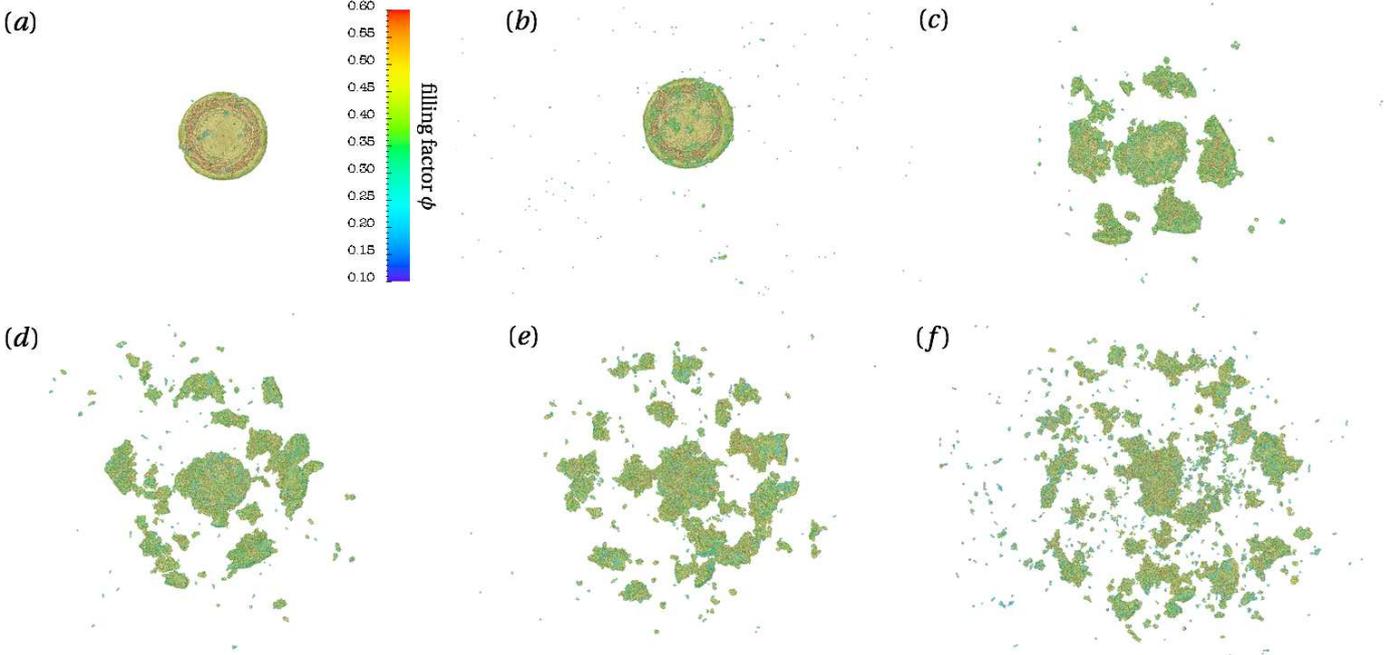}}
      \caption{Outcome of a collision between a target and projectile with the same $\phis$ for different inhomogeneities. In all cases, the target and projectile radii are $\rt = \unit[10]{cm}$ and $\rp = \unit[6]{cm}$, respectively, and the collision velocity is $v_0 = \unit[10]{\msec}$. Analogous to the initial targets in Fig.~\ref{fig:initial-setup}, the standard deviations for the Gaussian are $\phis = 0$ (a), $\phis = 0.01$ (b), $\phis= 0.02$ (c), $\phis = 0.03$ (d), $\phis = 0.04$ (e), and $\phis = 0.05$ (f). The collision outcome is shown in the impact direction. In the homogeneous case (a) and for small inhomogeneities (b), the target stays intact and forms one massive object with the projectile. For $\phis \ge 0.02$, the target fragments (c-d). The fragment sizes decrease with increasing $\phis$ and at the same time the number of fragments increases.}
         \label{fig:final-state}
\end{figure*}

We now perform test simulations using our inhomogeneity approach. The initial setup is the same as in most simulations of \citet{Geretshauser.2011}. It consists of two colliding spheres with $\phii = 0.35$. The radii of the target and projectile are $\rt = \unit[10]{cm}$ and $\rp = \unit[6]{cm}$, and they consist of 238,238 and 51,477 SPH particles, respectively, with masses of $\unit[1.23 \ten{-5}]{kg}$ each. The SPH particles are set on a cubic grid separated by a distance of \unit[2.6]{mm}. We consider two impact velocities of $v_0 = \unit[10]{\msec}$ and $v_0 = \unit[12.5]{\msec}$. For homogeneous aggregates, the smaller velocity leads to the sticking of the projectile to the target (see Fig.~\ref{fig:final-state}, a). In contrast, the higher velocity results in fragmentation \citep[see also][]{Geretshauser.2011}. For these two collision velocities, the inhomogeneity of both the target and the projectile is defined by the standard deviation of the Gaussian $\phis$, which is varied from 0 to 0.05 yielding the distributions depicted in Fig.~\ref{fig:initial-ff}. The resulting targets are shown in Fig.~\ref{fig:initial-setup}. The projectiles display a similar density pattern.

For $v_0 = \unit[10]{\msec}$, the final fragment distribution is shown in Fig.~\ref{fig:final-state} from the impact direction. For homogeneous aggregates (a) and the small standard deviation ($\phis = 0.01$, b), the target remains intact and the target and projectile form one large aggregate. For $\phis = 0.01$, some small fragments are visible. For $\phis \ge 0.02$, the target fragments and with greater inhomogeneity the largest fragments of the distribution decrease in size.

The first impression from our visual inspection of the collision products is confirmed by the more detailed analysis. For this purpose, we utilise the four-population model \citep{Geretshauser.2011}. According to this classification scheme of collision outcome, four classes of fragments are distinguished according to their mass: the largest fragment (index ``1'') and the second largest fragment (index ``2'') are characterised by single values of mass $m$, kinetic energy $E$, and other parameters. To account for the resolution limit, the sub-resolution population (index ``sr'') is introduced, which contains all fragments consisting of a single SPH particle (mass $m_{\rm SPH}$). This class is mainly described by global quantities. Fragments with masses $m_{\rm SPH} < m_{\rm frag} < m_2$ form a class whose cumulative mass distribution is described by a power-law (index ``pw''). This class is characterised by global quantities and distribution functions.

\begin{figure}
   \resizebox{\hsize}{!}
            {\includegraphics{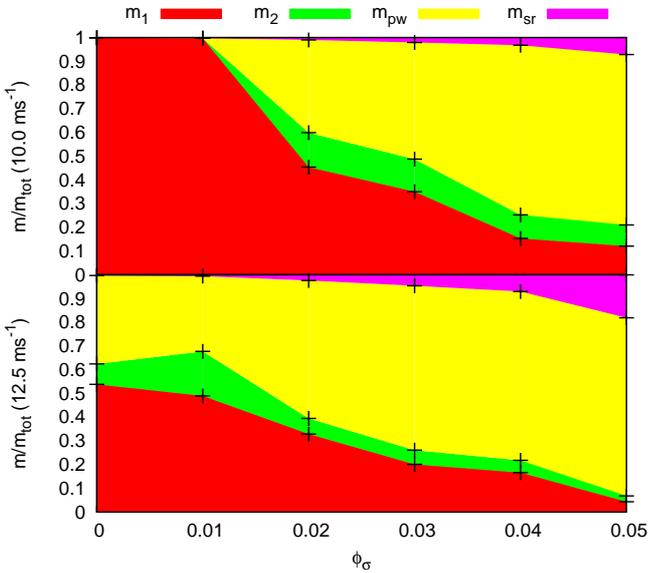}}
      \caption{The total masses of the four populations with inhomogeneities measured by $\phis$. The collision velocities are $v_0 =\unit[10]{\msec}$ (top) and $v_0 = \unit[12.5]{\msec}$ (bottom). The masses of the largest fragment $\mlst$, of the second largest fragment $\msnd$, of the power-law population $\mpwlw$, and of the sub-resolution population $\msr$ are stacked and normalised by the total mass of the system $m_{\rm tot}$. In both cases, $\mlst$ and $\msnd$ decrease with increasing $\phis$. Conversely, $\mpwlw$ and $\msr$ increase. For the homogeneous case ($\phis = 0$), sticking is found for the low velocity collision, whereas for $v_0 = \unit[12.5]{\msec}$ $\msnd \ne 0$ and $\mpwlw \ne 0$, which indicates fragmentation.}
         \label{fig:mass-fourpop}
\end{figure}

The evolution of the masses of each population with $\phis$ is shown in Fig.~\ref{fig:mass-fourpop} as a fraction of the total mass. The upper and the lower plots display the results for $v_0 = \unit[10]{\msec}$ and $v_0 = \unit[12.5]{\msec}$, respectively. The exact values can be found in Tab.~\ref{tab:mass-energy}. For the lower collision velocity and $\phis \le 0.01$, nearly all mass is stored in the largest fragment $\mlst$. For $\phis > 0.01$, a second largest fragment and a power-law population appears. With increasing inhomogeneity $\mlst$ rapidly decreases, while the mass of the second largest fragment $\msnd$ only slightly decreases. The masses $\mlst$ and $\msnd$ become comparable in the end. For high inhomogeneity values, most of the mass is stored in the power-law population $\mpwlw$, as we discuss below. However, for high inhomogeneities $\mpwlw$ seems to remain fairly constant, while the mass of the sub-resolution population $\msr$ increases. For the higher collision value (bottom plot of Fig.~\ref{fig:mass-fourpop}), the evolution is similar but starts with $\mpwlw \ne 0$. The mass $\mlst$ constantly decreases, but not as rapidly as in the low velocity case. Moreover, $\msnd$ at first increases, then slightly decreases until the largest and second largest fragment become comparable. The mass of the power-law population rapidly increases for $0.01 \le \phis \le 0.04$. For larger $\phis$, the power-law population only slightly increases, whereas $\msr$ increases at a higher rate. The mass evolution with increasing inhomogeneity can be summarised as follows: (1) inhomogeneity leads to fragmentation at collision velocities, at which homogeneous aggregates do not fragment, i.e.\ inhomogeneous aggregates are weaker.	 (2) For the two investigated velocities a larger $\phis$ leads to a decrease in the mass of the largest and second largest fragments. (3) A higher degree of inhomogeneity causes an increase in both the masses of the power-law and sub-resolution population.

\begin{figure}
   \resizebox{\hsize}{!}
            {\includegraphics{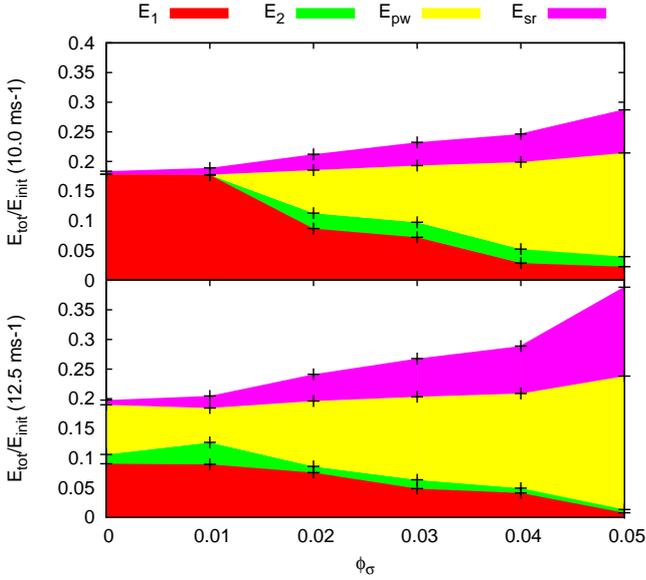}}
      \caption{The energy of each population for increasing inhomogeneity parameter $\phis$ and collision velocity $v_0 = \unit[10]{\msec}$ (top) and $v_0 = \unit[12.5]{\msec}$ (bottom). The energy of each population is the sum of translational, rotational and librational energy and normalised by the initial kinetic energy $E_{\rm init}$ . The two latter energy contributions are negligible for the head-on collisions presented here. Owing to the strong correlation with mass, this diagram follows a similar trend to Fig.~\ref{fig:mass-fourpop}. The energy contributions are divided up into $\Elst$, $\Esnd$, $\Epwlw$, and $\Esr$ for the largest and second largest fragment, the power-law and sub-resolution population, respectively. In both velocity cases, both $\Elst$ and $\Esnd$ decrease with increasing $\phis$. In contrast, $\Epwlw$ and $\Esr$ increase. In particular, the high fraction of kinetic energy stored in the sub-resolution population is remarkable.}
         \label{fig:final-energy}
\end{figure}

We turn to the analysis of the distribution of the residual total energy after the collision, which is shown in Fig.~\ref{fig:final-energy} as a fraction of the initial kinetic energy. The residual total energy is the sum of the translational energy and the energy stored in the internal degrees of freedom (e.g.\ rotation, libration) of the respective populations. The tabulated values can be found in Tab.~\ref{tab:mass-energy}. In Fig.~\ref{fig:final-energy}, the low velocity case with $v_0 = \unit[10]{\msec}$ is shown in the upper plot and the high velocity case with $v_0 = \unit[12.5]{\msec}$ in the lower plot. For the former, a large amount of energy $\Elst$ is stored in the largest fragment, but a considerable amount $\Esr$ is also stored in the sub-resolution population. As $\Elst$ decreases, the contribution of the second largest fragment at first increases, then decreases slightly with larger $\phis$. As for the mass, the energy contribution of the power-law population $\Epwlw$ increases with a higher degree of inhomogeneity. Remarkably, the total energy fraction of these three populations remains almost constant. In contrast, the energy contribution of the sub-resolution population $\Esr$ increases more strongly than its mass contribution in Fig.~\ref{fig:mass-fourpop}. The overall residual energy increases with a higher degree of inhomogeneity and this energy excess is stored mostly in the sub-resolution population. For the high velocity case (bottom plot of Fig.~\ref{fig:final-energy}), this is even more evident: $\Elst$, $\Esnd$, and $\Epwlw$ behave similar to the low velocity case but with $\Esnd \ne 0$ and $\Epwlw \ne 0$ for $\phis = 0$. For a higher degree of inhomogeneity, less energy can again be dissipated and the total residual energy of the largest three populations remains nearly constant for increasing $\phis$. As a consequence, the residual energy excess is mainly stored in the sub-resolution population for large values of $\phis$. The conclusions of our energy analysis are: (1) with increasing impact velocity less energy can be dissipated by the system and the residual energy increases as previously found by \citet{Geretshauser.2011}. (2) With increasing inhomogeneity, the effectiveness of the energy dissipation decreases for the two investigated velocities. (3) The energy contribution of all populations show a similar dependence on $\phis$ to their mass contributions: the fractions of energy stored in the largest and second largest fragment follow a decreasing trend with increasing inhomogeneity. The contribution of power-law and sub-resolution populations increase. (4) While the overall energy contribution of the largest fragment populations remains nearly constant with $\phis$, the fraction stored in the sub-resolution population drastically increases, such that the energy that cannot be dissipated owing to the increasing inhomogeneity is stored in rapidly moving single SPH particles.

\begin{figure}
   \resizebox{\hsize}{!}
            {\includegraphics{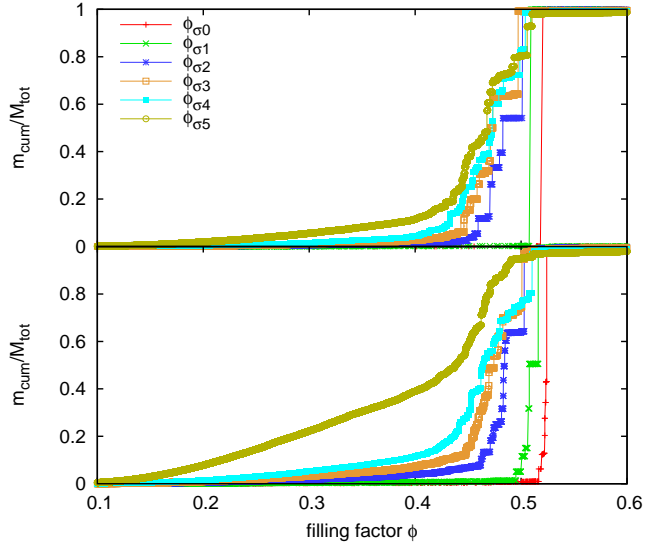}}
      \caption{Cumulative mass of the fragment masses over their average filling factor \protect{$\phi$}. The mass is normalised by the total fragment mass $\mtot$. The collision velocities are $v_0 = \unit[10]{\msec}$ (top) and $v_0 = \unit[12.4]{\msec}$ (bottom). Compared to the initial distribution of filling factors (Fig.\ \ref{fig:initial-ff}), the curves are shifted significantly towards higher filling factors. This reflects the compression taking place during the collision.}
         \label{fig:final-ff}
\end{figure}

In comparison to the initial filling factor distribution of Fig.~\ref{fig:initial-ff}, the final filling factor distribution of the cumulated mass is shifted towards higher filling factors. Figure \ref{fig:final-ff} reveals that in most cases more than $\unit[85]{\%}$ of the fragment mass is stored in filling factors $\gtrsim \unit[42]{\%}$ (except for $v_0 = \unit[12.5]{\msec}$ and $\phis = 0.050$). With the same exception, the distributions look quite similar in both velocity cases. There is a slight tendency for there to be more mass at lower filling factors in the collision with $v_0 = \unit[12.5]{\msec}$. From Fig.\ \ref{fig:final-ff}, it can also be seen that the broader the initial filling factor distribution, the broader the final filling factor distribution. For the $v_0 = \unit[10.0]{\msec}$ case, the two most homogeneous aggregates hardly produce any fragments that cause a jump in the curves without any intermediate steps. The distribution for the highest collision velocity ($v_0 = \unit[12.5]{\msec}$) and the highest degree of inhomogeneity ($\phis = 0.050$) is particularly interesting since $\sim \unit[30]{\%}$ of the fragment mass possess filling factors $< 0.35$, which is a larger fraction than for the initial aggregate. This implies that rarefaction occurs during the collision. On the microscopic scale,  this is equivalent to monomer re-ordering and stretching of monomer chains. We note that a tiny fraction of particles acquire overdensities ($\phi > 1$) in the collision. In the most violent case, this applies to single SPH particles, which in total represent \unit[0.03]{\%} of the total mass of the system, which is negligible. These overdensities indicate that the monomers are breaking up on the microscopic scale.

\begin{figure}
   \resizebox{\hsize}{!}
            {\includegraphics{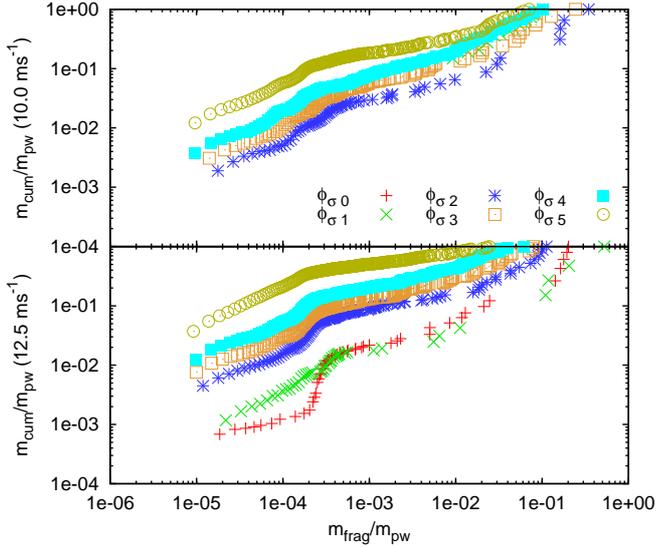}}
      \caption{Cumulative mass distribution of a power-law population for different inhomogeneity parameters $\phis$ and collision velocities $v_0 = \unit[10]{\msec}$ (top) $v_0 = \unit[12.5]{\msec}$ (bottom). In both velocity cases both the mass of the largest member of the power-law population and the power-law slope decrease. This indicates the fragmentation to smaller aggregates for increasing $\phis$. For equal $\phis$, the slopes are shallower for the higher velocity. The power-law fit parameters can be found in Tab.~\ref{tab:pwlw-fits}.}
         \label{fig:mass-pwlw}
\end{figure}

\begin{figure}
   \resizebox{\hsize}{!}
            {\includegraphics{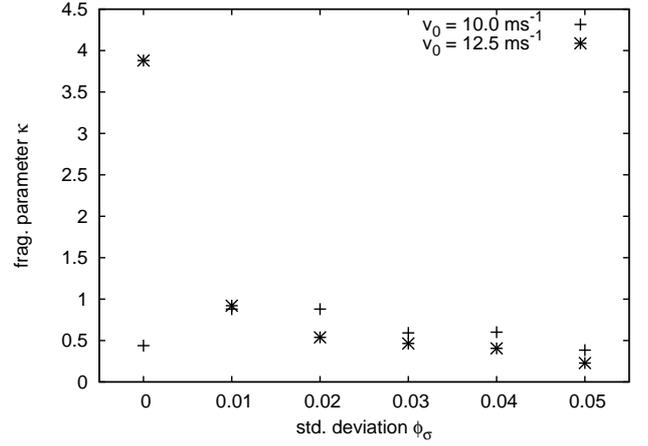}}
      \caption{Fragmentation parameter $\kappa$ of the power-law population. For increasing inhomogeneity, indicated by the standard deviation $\phis$, the fragmentation parameter decreases. This indicates shallower slopes of the power-law mass distribution which is interpreted as an increasing fraction of smaller fragments. For $\phis \ge 0.02$, the higher impact velocity results in smaller values of $\kappa$.}
         \label{fig:kappa-pwlw}
\end{figure}

We have so far considered all members of the four-population model. We now constrain our analysis to the power-law population. Its mass distribution is shown in the cumulative plot of Fig.~\ref{fig:mass-pwlw} again for $v_0 = \unit[10]{\msec}$ (top) and $v_0 = \unit[12.5]{\msec}$ (bottom). The figure shows the ratio of the cumulative mass $\mcum$ to the fragment mass $\mf$, which are both normalised by the total mass of the power-law population $\mpwlw$ given in Tab.~\ref{tab:mass-energy}. The cumulative distribution can be accurately \citep[e.g.][]{Guttler.2010} described by a power law
\begin{equation}
\mcum (\mf) = \int^{\mf}_0 n(m) \, m \, \rmd m = \left( \frac{\mf}{\mupwlw} \right)^\kappa \; ,
\label{eq:cum-pwlw}
\end{equation}
where $n(m) \, \rmd m$ is the number density of fragments of mass $m$, which is normalised by the total mass of the power-law population $\mpwlw$. The quantity $\mupwlw$ is the normalised mass of the most massive member of the power-law population and $\kappa$ is the power-law index or fragmentation parameter. We fitted the cumulative mass distributions of Fig.~\ref{fig:mass-pwlw} for all fragment masses with Eq.~\ref{eq:cum-pwlw}. Finding an explanation for the kink in the cumulative mass curves is the subject of ongoing research. This deviation from a power law might be either a physical or a resolution effect \citep[see also][]{Geretshauser.2011}. The results are listed in Tab.~\ref{tab:pwlw-fits}. Figure \ref{fig:kappa-pwlw} shows that the power-law index $\kappa$ decreases slightly as the degree of inhomogeneity increases for both collision velocities, which leads to increasingly shallower slopes in Fig.~\ref{fig:mass-pwlw} owing to the production of smaller fragments with increasing $\phis$. In addition, the $\kappa$ values for the higher impact velocity ($v_0 = \unit[12.5]{\msec}$) are systematically below the $\kappa$ values of the lower velocity ($v_0 = \unit[10.0]{\msec}$), except for small values of $\phis$ for which there is only a small number of fragments (see Tab.~\ref{tab:mass-energy}) and hence the quality of the statistics is insufficient. This finding indicates that a larger fraction of less massive fragments are produced at higher impact velocities, which is expected. 

\begin{figure}
   \resizebox{\hsize}{!}
            {\includegraphics{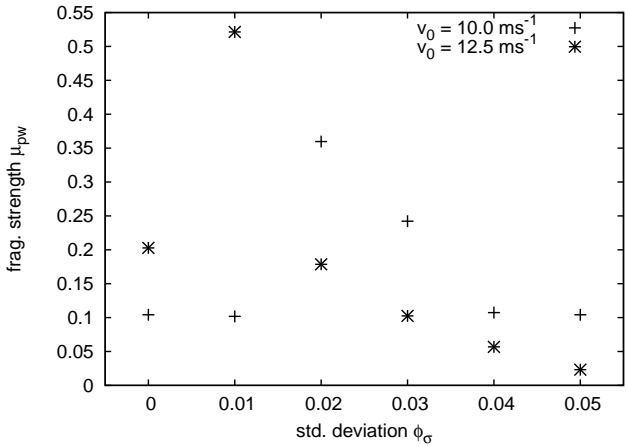}}
      \caption{Fragmentation strength of the power-law population. The quantity $\mupwlw$ represents the normalised mass of the most massive member of the power-law population and serves as an indicator of the fragmentation strength. For increasing inhomogeneity, represented by larger values of $\phis$, $\mupwlw$ at first increases and then decreases again. The increase may be caused by our low quality statistics owing to the small number of fragments.}
         \label{fig:mu-pwlw}
\end{figure}

As shown in Fig.~\ref{fig:mu-pwlw}, the mass of the largest member of the power-law population, given by $\mupwlw$, at first increases for larger values of $\phis$. It has a maximum at $\phis \sim 0.02$ for $v_0 = \unit[10]{\msec}$ and at $\phis \sim 0.01$ for $v_0 = \unit[12.5]{\msec}$, and then decreases in an exponential fashion. When comparing both velocity cases (see Tab.~\ref{tab:mass-energy}), we find that more fragments  are produced in the high velocity case. This was also observed by \citet{Geretshauser.2011}. The influence of the inhomogeneity on the power-law population can be summarised as follows: (1) the overall mass of the power-law distribution is increased as discussed above; (2) a higher degree of inhomogeneity leads to the production of smaller fragments; (3) the largest member of the power-law population reaches its highest mass at small $\phis$, after which its mass decreases with increasing $\phis$.

\begin{table*}
\centering
\begin{tabular*}{\hsize}{c@{\extracolsep{\fill}}c c c c c c c c c c c c}
\hline\hline
\input{mass-energy-tab.tex}
\hline
\end{tabular*}
\caption{Simulation results presented according to the four-population classification. The subscripts ``1'', ``2'', ``pw'', ``sr'', and ``tot'' denote the largest and second largest fragment, the power-law, sub-resolution, and total fragment population, respectively. The quantities $E$, $m$, and $N$ are the total energy, the mass, and the number of fragments, respectively. The standard deviation for the inhomogeneity Gaussian distribution is given by $\phis$ and the impact velocity by $v_0$.}
\label{tab:mass-energy}
\end{table*}

\begin{table}
\centering
\begin{tabular*}{\hsize}{c@{\extracolsep{\fill}}c c c}
\hline\hline
\input{pwlw-fits.tex} \\
\hline
\end{tabular*}
\caption{Fit values of the power-law population.The quantities $\slpwlw$ and $\mupwlw$ denote, respectively, the slope of the power-law fit and the mass of the most massive member, which is normalised by the total mass of the power-law population $\mpwlw$. The standard deviation of the inhomogeneity Gaussian is denoted by $\phis$ and the impact velocity by $v_0$.}
\label{tab:pwlw-fits}
\end{table}


\section{Discussion and outlook}
\label{sec:discussion}

We have presented the first simulations of inhomogeneous porous pre-planetesimals, where the inhomogeneity of \sio dust aggregates is explicitly resolved. This inhomogeneity can in principal also be determined in laboratory experiments \citep{Guttler.2009}. The approach is based on the concept that according to our porosity model \citep{Geretshauser.2010} inhomogeneities in filling factor cause fluctuations in the compressive, shear, and tensile strength in the aggregate. These fluctuations can be regarded as flaws in the material. In contrast to damage models designed for brittle material \citep{Grady.1980}, we did not explicitly evolve the propagation of these flaws. Instead, the defects in the dust material, which behaves more like a fluid, were approximated by the time evolution of the filling factor or - equivalently - the density. The inhomogeneity of an aggregate was assumed in the initial SPH particle distribution by using a Gaussian distribution to model the filling factor. The inhomogeneity is characterised by the standard deviation $\phis$ of the Gaussian and the typical radius $R_{\rm c}$ of a clump.

Using this inhomogeneity approach, we have performed test simulations of collisions between dust aggregates of intermediate porosity with fixed typical clump size. Two collision velocities have been chosen: one below and one above the fragmentation threshold for homogeneous aggregates. The results have been analysed using the four-population model of \citet{Geretshauser.2011}. For the lower collision velocity, we have shown that inhomogeneity leads to fragmentation. For both velocities, the masses of largest and second largest fragment are lower for a higher degree of inhomogeneity whereas the masses of the power-law and sub-resolution population are higher. Focussing on the power-law population, the number of fragments and the fraction of small fragments increase with increasing $\phis$. These findings show that the inhomogeneity approach has passed the first test in explicitly resolving the heterogeneity of porous objects. In a future study, we will address the influence of the typical clump size on the outcome of pre-planetesimal collisions.

Our findings indicate that inhomogeneous dust aggregates are weaker than their homogeneous equivalents. A slight inhomogeneity is sufficient to cause a catastrophic disruption instead of growth as the result of a dust aggregate collision. Therefore, inhomogeneity might explain the higher velocity thresholds for fragmentation in simulations than the a lower value found in laboratory experiments \citep{Geretshauser.2011}.

Furthermore, macroscopic dust aggregates in protoplanetary discs are produced by the subsequent impacts of smaller aggregates at different impact velocities, i.e.\ pre-planetesimals have a collision history, thus are very likely to be inhomogeneous. This has a negative effect on planetesimal formation by subsequent collisions because pre-planetesimals might become weaker as the  number of collisions increases. As a result, their threshold velocity for catastrophic disruption might be smaller. However, in a future study we will investigate the influence of the clump size of the stability of the pre-planetesimals. If the clumps are small relative to the pre-planetesimal size, this might increase the stability of the aggregates. 

With the model presented in this paper, we are able to include these features. Further studies might be carried out that investigate the inhomogeneity created by subsequent multiple impacts. The result could be classified according to the inhomogeneity model that we have presented here. 

To date it has been found that the simulations of collisions between homogeneous aggregates carried out by means of the porosity model by \citet{Geretshauser.2010} depend only on the filling factor and produce the same fragment distribution for an identical set of input parameters. By randomly assigning an inhomogeneity pattern it is now possible to profoundly investigate the statistics of a fragment distribution. Statistical fluctuations in the quantities of the four-population model can now be estimated in particular for simulations with a small number of fragments.

In high velocity grazing collisions, the target begins to rotate. For highly porous aggregates, which have a low tensile strength, high spinning rates tend to lead to fragmentation of the target. With increasing inhomogeneity, this might also be true for aggregates with medium and low porosity. A quantitative investigation of this effect can be carried out by means of the presented approach.

The filling factor distribution of dust aggregates can be determined in the laboratory by X-ray tomography measurements \citep{Guttler.2009}. These empirical data can be directly implemented into our inhomogeneity model. In addition the successful material calibration \citep{Geretshauser.2010}, the obtained inhomogeneity parameters can be used to improve the realistic simulation of porous dust aggregates.

\begin{acknowledgements}
R.J.G. and R.S. wish to thank S.\ Arena, W.\ Benz, and F.\ Meru for many illuminating discussions. We are obliged to our referee Gareth Collins for his profound revision and constructive comments which helped to improve and clarify the manuscript. The SPH simulations were performed on the university cluster of the computing centre of T\"ubingen, the bwGriD clusters in Karlsruhe, Stuttgart, and T\"ubingen. Computing time was also provided by the High Performance Computing Centre Stuttgart (HLRS) on the national supercomputer NEC Nehalem Cluster under project grant SPH-PPC/12848. This project was funded by the Deutsche Forschungsgemeinschaft within the Forschergruppe ``The Formation of Planets: The Critical First Growth Phase" under grant Kl 650/8-1.
\end{acknowledgements}

\begin{appendix}

\section{List of symbols}

\begin{table}
\caption{List of symbols}             
\label{tab:symbol-list}      
\centering                          
\begin{tabular*}{\hsize}{l@{\extracolsep{\fill}}l}        
\hline\hline                 
symbol & explanation \\    
\hline                        
$c_{\rm g}$		& crack propagation speed \\
$D$				& damage parameter\\
$\Elst$, $\Esnd$, $\Epwlw$, $\Esr$ & total energies of largest and second largest \\
				&  fragment, power-law and sub-res.\ populations \\
$\Etot$			& total energy of the fragment population \\
$h$				& smoothing length \\
$J_2$			& second irreducible invariant\\
$k_{\rm wb}$		& number density of flaws in the Weibull distribution \\
$K(\phi)$			& bulk modulus \\
$K_0$			& bulk modulus for the uncompressed dust sample \\
$m$				& mass \\
$\mlst$, $\msnd$, $\mpwlw$, $\msr$ & masses of largest and second largest fragment, \\
				& power-law and sub-resolution populations \\
$\mcum$			& cumulated mass \\
$m_{\rm frag}$		& fragment mass \\
$\mproj$			& projectile mass \\
$\mtar$			& target mass \\
$m_{\rm wb}$		& exponent in the Weibull distribution \\
$n(\phi)$			& number density of volumes with filling factor $\phi$ \\
$n(m)$			& number density of fragments with mass $m$ \\
$n_{\rm act}$		& number of activated flaws \\
$n_{\rm frag}$		& number of fragments \\
$\Npwlw$, $\Nsr$	& number of fragments in the power-law \\ 
				& and sub-resolution populations \\
$p$				& hydrostatic pressure \\
$\pmean$			& mean pressure of $\Sigma$ \\
$\rproj$, $\rtar$		& projectile and target radii \\
$R_{\rm c}	$		& typical clump size \\		
$R_{\rm s}	$		& sphere of crack evolution \\		
$\Sab$			& deviatoric stress tensor \\
$t$				& time \\
$T(\phi)$			& tensile strength \\
$v_0$			& relative collision velocity \\
$\va$			& velocity \\
$W(\vx^a, \vx^b, h)$	& smoothing kernel function \\
$\xa$			& position \\	
$Y(\phi)$			& shear strength \\

$\gamma$		& exponent of $K(\phi)$ \\
$\delta_{\alpha\beta}$& Kronecker symbol \\
$\Delta$			& power parameter of $\Sigma$ \\
$\epsilon_{\rm act}$	& strain activation threshold \\
$\eab$			& strain tensor \\
$\kappa$			& fragmentation parameter \\
$\mu (\phi)$		& shear modulus \\
$\mupwlw$		& normalised mass of the most massive\\
				& member of the power-law population \\
$\rho$			& density \\
$\rhos$			& matrix density \\
$\stab$			& stress tensor \\
$\phi$			& filling factor \\
$\phi_0'$			& reference filling factor at $p=0$ \\
$\phi_c^+$, $\phi_c^-$	& threshold filling factors from elastic\\
				& to plastic compression/tension \\
$\phi_{\rm max}$, $\phi_{\rm min}$	& maximum/minimum filling factors of $\Sigma$ \\
$\phi_{\rm i}$		& initial filling factor \\
$\phi_{\rm \mu}$	& inhomogeneity median \\
$\phis$			& inhomogeneity standard deviation \\
$\phirbd$			& filling factor of an RBD dust sample \\
$\Sigma(\phi)$		& compressive strength \\

\hline                                   
\end{tabular*}
\end{table}

\end{appendix}

\bibliographystyle{aa}
\bibliography{literature}

\end{document}

%% file: definedquantities.tex
\newcommand{\totder}[2]{\frac{{\rm d} #1}{{\rm d} #2}}
\newcommand{\partder}[2]{\frac{\partial #1}{\partial #2}}

\newcommand{\vx}{\mathbf{x}}					
\newcommand{\xa}{x_\alpha}					
\newcommand{\xb}{x_\beta}					
\newcommand{\va}{v_\alpha}					
\newcommand{\dab}{\delta_{\alpha\beta}}			

\newcommand{\stab}{\sigma_{\alpha\beta}}		

\newcommand{\rmd}{\mathrm{d}}				
\newcommand{\eab}{\epsilon_{\alpha\beta}}		
\newcommand{\Sab}{S_{\alpha\beta}}			
\newcommand{\rhos}{\rho_{\rm s}}				
\newcommand{\phii}{\phi_{\rm i}}				
\newcommand{\pmean}{p_{\rm m}}				

\newcommand{\sio}{$\mathrm{SiO_2}$~}			
\newcommand{\phimax}{\phi_{\rm max}}			
\newcommand{\phimin}{\phi_{\rm min}}			
\newcommand{\phirbd}{\phi_{\rm RBD}}			

\newcommand{\rt}{r_{\rm t}}					
\newcommand{\rtar}{r_{\rm t}}					
\newcommand{\rp}{r_{\rm p}}					
\newcommand{\rproj}{r_{\rm p}}					

\newcommand{\mtot}{m_{\rm tot}}				
\newcommand{\mlst}{m_{1}}					
\newcommand{\msnd}{m_{2}}					
\newcommand{\mpwlw}{m_{\rm pw}}				
\newcommand{\msr}{m_{\rm sr}}				

\newcommand{\msec}{\rm ms^{-1}}
\newcommand{\mtar}{m_{\rm t}}
\newcommand{\mproj}{m_{\rm p}}

\newcommand{\phis}{\phi_{\sigma}}				
\newcommand{\mcum}{m_{\rm cum}}				
\newcommand{\mf}{m_{\rm frag}}					
\newcommand{\slpwlw}{\kappa}				
\newcommand{\mupwlw}{\mu_{\rm pw}}			
\newcommand{\Etot}{E_{\rm tot}}				
\newcommand{\Elst}{E_{1}}					
\newcommand{\Esnd}{E_{2}}					
\newcommand{\Epwlw}{E_{\rm pw}}				
\newcommand{\Esr}{E_{\rm sr}}					
\newcommand{\Npwlw}{N_{\rm pw}}				
\newcommand{\Nsr}{N_{\rm sr}}					
\newcommand{\ten}[1]{\times 10^{#1}}


%% file: mass-energy-tab.tex
$\phis$ & $v_0$ & $\Etot$ & $\mlst$ & $\Elst$ & $\msnd$ & $\Esnd$ & $\mpwlw$ & $\Epwlw$ & $\Npwlw$ & $\msr$ & $\Esr$ & $\Nsr$ \\
	& [$\msec$]	& [J]	& [kg] & [J] & [kg] & [J] & [kg] & [J] & & [kg] & [J] &
\\ \hline \\
0.00	& 10.0 & 5.81 & 3.56 & 5.64 & $5 \ten{-5}$ & $1 \ten{-4}$	& $5 \ten{-3}$	& $3 \ten{-3}$	& 17 & $2 \ten{-3}$ & 0.17 & 158 \\ 
0.01 & 10.0 & 5.99 & 3.55 & 5.61 & $3 \ten{-4}$ & $7 \ten{-4}$	& $3 \ten{-3}$	& 0.01 		& 36 & 0.01 & 0.36 & 610 \\ 
0.02 & 10.0 & 6.72 & 1.62 & 2.75 & 0.52 		& 0.82 		& 1.39		& 2.30 		& 290 & 0.04 & 0.84 & 2854 \\ 
0.03 & 10.0 & 7.35 & 1.25 & 2.28 & 0.49 		& 0.80 		& 1.75 		& 3.03 		& 588 & 0.07 & 1.24 & 6043 \\ 
0.04 & 10.0 & 7.80 & 0.55 & 0.91 & 0.35 		& 0.75 		& 2.55 		& 4.64 		& 1057 & 0.11 & 1.50 & 9292 \\ 
0.05 & 10.0 & 9.09 & 0.43 & 0.71 & 0.32 		& 0.54 		& 2.56 		& 5.54 		& 3056 & 0.25 & 2.30 & 20657 \\ 
\hline \\
0.00 & 12.5 & 9.78 & 1.92 & 4.47 & 0.31 		& 0.78 		& 1.33 		& 4.14 		& 123 & 0.01 & 0.39 & 861 \\ 
0.01 & 12.5 & 10.12 & 1.74 & 4.42 & 0.67 		& 1.83 		& 1.13 		& 2.87 		& 157 & 0.02 & 1.00 & 1650 \\ 
0.02 & 12.5 & 11.93 & 1.17 & 3.73 & 0.23 		& 0.52 		& 2.07 		& 5.45 		& 984 & 0.08 & 2.23 & 6882 \\ 
0.03 & 12.5 & 13.24 & 0.72 & 2.40 & 0.21 		& 0.74 		& 2.47 		& 6.91 		& 1865 & 0.16 & 3.19 & 13285 \\ 
0.04 & 12.5 & 14.29 & 0.59 & 2.03 & 0.18 		& 0.40 		& 2.54 		& 7.90 		& 3180 & 0.25 & 3.96 & 20402 \\ 
0.05 & 12.5 & 19.20 & 0.15 & 0.38 & 0.09 		& 0.27 		& 2.68 		& 11.13 		& 10578 & 0.65 & 7.42 & 52478 \\ 

%% file: pwlw-fits.tex
$\phis$	&	$v_0$	&	$\slpwlw$			& 	$\mupwlw$ \\
\hline \\
0		& 10.0	& $0.44	\pm 0.027$		& $0.10	\pm 2.4 \ten{-3}$ \\
0.01		& 10.0	& $0.88	\pm 0.035$		& $0.10	\pm 2.3 \ten{-3}$ \\
0.02		& 10.0	& $0.88	\pm 0.046$		& $0.36	\pm 0.012$ \\
0.03		& 10.0	& $0.59	\pm 0.011$		& $0.24	\pm 6.9 \ten{-3}$ \\
0.04		& 10.0	& $0.60	\pm 6.4 \ten{-3}$	& $0.11	\pm 1.7 \ten{-3}$ \\
0.05		& 10.0	& $0.38	\pm 5.4 \ten{-3}$	& $40.10	\pm 4.0 \ten{-3}$ \\
\hline \\
0		& 12.5	& $3.88		\pm 0.254$	& $0.20	\pm 1.4 \ten{-3}$ \\
0.01		& 12.5	& $0.92		\pm 0.030$	& $0.52	\pm 1.0 \ten{-2}$ \\
0.02		& 12.5	& $0.54		\pm 0.017$	& $0.18	\pm 1.0 \ten{-2}$ \\
0.03		& 12.5	& $0.46		\pm 0.006$	& $0.10	\pm 3.0 \ten{-3}$ \\
0.04		& 12.5	& $0.41		\pm 0.004$	& $0.06	\pm 1.3 \ten{-3}$ \\
0.05		& 12.5	& $0.23		\pm 0.003$	& $0.02	\pm 1.0 \ten{-4}$

%% file: damagepaper_arxiv.bbl
\begin{thebibliography}{34}
\expandafter\ifx\csname natexlab\endcsname\relax\def\natexlab#1{#1}\fi

\bibitem[{Benz(2000)}]{Benz.2000}
Benz, W. 2000, Space Science Reviews, 92, 279

\bibitem[{Benz \& Asphaug(1994)}]{Benz.1994}
Benz, W. \& Asphaug, E. 1994, Icarus, 107, 98

\bibitem[{Benz \& Asphaug(1995)}]{Benz.1995}
Benz, W. \& Asphaug, E. 1995, Computer Physics Communications, 87, 253

\bibitem[{Birnstiel {et~al.}(2010)Birnstiel, Ricci, Trotta, Dullemond, Natta,
  Testi, Dominik, Henning, Ormel, \& Zsom}]{Birnstiel.2010}
Birnstiel, T., Ricci, L., Trotta, F., {et~al.} 2010, Astronomy and
  Astrophysics, 516, L14

\bibitem[{Blum \& Schr\"apler(2004)}]{Blum.2004}
Blum, J. \& Schr\"apler, R. 2004, Phys. Rev. Lett., 93, 115503

\bibitem[{Blum \& Wurm(2008)}]{Blum.2008}
Blum, J. \& Wurm, G. 2008, Annual Review of Astronomy and Astrophysics, 46, 21

\bibitem[{Brauer {et~al.}(2008)Brauer, Dullemond, \& Henning}]{Brauer.2008}
Brauer, F., Dullemond, C.~P., \& Henning, T. 2008, Astronomy and Astrophysics,
  480, 859

\bibitem[{Dullemond \& Dominik(2005)}]{Dullemond.2005}
Dullemond, C.~P. \& Dominik, C. 2005, Astronomy and Astrophysics, 434, 971

\bibitem[{Dullemond {et~al.}(2007)Dullemond, Hollenbach, Kamp, \&
  D'Alessio}]{Dullemond.2007}
Dullemond, C.~P., Hollenbach, D., Kamp, I., \& D'Alessio, P. 2007, in
  {P}rotostars and {P}lanets {V}, ed. B.~Reipurth, D.~Jewitt, \& K.~Keil
  (Tucson: University of Arizona Press), 555--572

\bibitem[{Geretshauser {et~al.}(2011)Geretshauser, Meru, Speith, \&
  Kley}]{Geretshauser.2011}
Geretshauser, R., Meru, F., Speith, R., \& Kley, W. 2011, Astronomy and
  Astrophysics, 531, A166

\bibitem[{Geretshauser {et~al.}(2010)Geretshauser, Speith, G\"uttler, Krause,
  \& Blum}]{Geretshauser.2010}
Geretshauser, R.~J., Speith, R., G\"uttler, C., Krause, M., \& Blum, J. 2010,
  Astronomy and Astrophysics, 513, A58

\bibitem[{Gingold \& Monaghan(1977)}]{Gingold.1977}
Gingold, R.~A. \& Monaghan, J.~J. 1977, Monthly Notices of the Royal
  Astronomical Society, 181, 375

\bibitem[{Grady \& Kipp(1980)}]{Grady.1980}
Grady, D.~E. \& Kipp, M.~E. 1980, International Journal of Rock Mechanics and
  Mining Sciences \& Geomechanics Abstracts, 17, 147

\bibitem[{G\"uttler {et~al.}(2010)G\"uttler, Blum, Zsom, Ormel, \&
  Dullemond}]{Guttler.2010}
G\"uttler, C., Blum, J., Zsom, A., Ormel, C.~W., \& Dullemond, C.~P. 2010,
  Astronomy and Astrophysics, 513, A56

\bibitem[{G\"uttler {et~al.}(2009)G\"uttler, Krause, Geretshauser, Speith, \&
  Blum}]{Guttler.2009}
G\"uttler, C., Krause, M., Geretshauser, R.~J., Speith, R., \& Blum, J. 2009,
  The Astrophysical Journal, 701, 130

\bibitem[{Herrmann(1969)}]{Herrmann.1969}
Herrmann, W. 1969, Journal of Applied Physics, 40, 2490

\bibitem[{Hipp \& Rosenstiel(2004)}]{Hipp.2004}
Hipp, M. \& Rosenstiel, W. 2004, in Lecture Notes in Computer Science, Vol.
  3149, Euro-Par, ed. M.~Danelutto, M.~Vanneschi, \& D.~Laforenza (Springer),
  189--197

\bibitem[{Jutzi {et~al.}(2008)Jutzi, Benz, \& Michel}]{Jutzi.2008}
Jutzi, M., Benz, W., \& Michel, P. 2008, Icarus, 198, 242

\bibitem[{Jutzi {et~al.}(2009)Jutzi, Michel, Hiraoka, Nakamura, \&
  Benz}]{Jutzi.2009}
Jutzi, M., Michel, P., Hiraoka, K., Nakamura, A.~M., \& Benz, W. 2009, Icarus,
  201, 802

\bibitem[{Libersky \& Petschek(1991)}]{Libersky.1991}
Libersky, L.~D. \& Petschek, A.~G. 1991, in Lecture notes in physics, Vol. 395,
  {A}dvances in the {F}ree-{L}agrange method: including contributions on
  adaptive gridding and the smooth particle hydrodynamics method, ed.
  H.~Trease, M.~J. Fritts, \& W.~P. Crowley (Springer)

\bibitem[{Libersky {et~al.}(1993)Libersky, Petschek, Carney, Hipp, \&
  Allahdadi}]{Libersky.1993}
Libersky, L.~D., Petschek, A.~G., Carney, T.~C., Hipp, J.~R., \& Allahdadi,
  F.~A. 1993, Journal of Computational Physics, 109, 67

\bibitem[{Lissauer \& Stewart(1993)}]{Lissauer.1993}
Lissauer, J.~J. \& Stewart, G.~R. 1993, in {P}rotostars and planets, ed. E.~H.
  Levy \& J.~I. Lunine, Space Science Series (Univ. of Arizona Press),
  1061--1088

\bibitem[{Lucy(1977)}]{Lucy.1977}
Lucy, L.~B. 1977, The Astronomical Journal, 82, 1013

\bibitem[{Monaghan(2005)}]{Monaghan.2005}
Monaghan, J.~J. 2005, Reports on Progress in Physics, 68, 1703

\bibitem[{Monaghan \& Lattanzio(1985)}]{Monaghan.1985}
Monaghan, J.~J. \& Lattanzio. 1985, Astronomy and Astrophysics, 149, 135

\bibitem[{Randles \& Libersky(1996)}]{Randles.1996}
Randles, P.~W. \& Libersky, L.~D. 1996, Computer Methods in Applied Mechanics
  and Engineering, 139, 375

\bibitem[{Sch\"afer(2005)}]{Schafer.2005}
Sch\"afer, C. 2005, PhD thesis, {Universit\"at T\"ubingen}, T\"ubingen

\bibitem[{Sch\"afer {et~al.}(2007)Sch\"afer, Speith, \& Kley}]{Schafer.2007}
Sch\"afer, C., Speith, R., \& Kley, W. 2007, Astronomy and Astrophysics, 470,
  733

\bibitem[{Sirono(2004)}]{Sirono.2004}
Sirono, S. 2004, Icarus, 167, 431

\bibitem[{Weibull(1939)}]{Weibull.1939}
Weibull, W. 1939, {I}ngeni\"orsvetenskapsakademiens handlingar, Vol. 151, {A}
  statistical theory of the strength of materials (Stockholm: Generalstabens
  litografiska anstalts f\"orlag)

\bibitem[{Weidenschilling(1977)}]{Weidenschilling.1977}
Weidenschilling, S.~J. 1977, Monthly Notices of the Royal Astronomical Society,
  180, 57

\bibitem[{Weidenschilling \& Cuzzi(1993)}]{Weidenschilling.1993}
Weidenschilling, S.~J. \& Cuzzi, J.~N. 1993, in {P}rotostars and planets, ed.
  E.~H. Levy \& J.~I. Lunine, Space Science Series (Univ. of Arizona Press),
  1031--1060

\bibitem[{Wilkins(1964)}]{Wilkins.1964}
Wilkins, M. 1964, in {M}ethods in {C}omputational {P}hysics V.3, ed. B.~Alder,
  S.~Fernbach, \& M.~Rotenberg (New York: Academic Press), 211--263

\bibitem[{Zsom {et~al.}(2010)Zsom, Ormel, {G\"uttler C.}, Blum, \&
  Dullemond}]{Zsom.2010}
Zsom, A., Ormel, C.~W., {G\"uttler C.}, Blum, J., \& Dullemond, C.~P. 2010,
  Astronomy and Astrophysics, 513, A57

\end{thebibliography}
